\documentclass{epl}
\usepackage{times}
\usepackage{amssymb}
\usepackage{version}
\usepackage{amsmath}
\usepackage{eufrak}
\renewcommand{\epsilon}{\varepsilon}
\renewcommand{\vec}{{}}
\newcommand{\E}{\varepsilon}
\newcommand{\rme}{\mathrm{e}}

\newcommand{\Tr}{{\mathrm{Tr}}}
\newcommand{\nn}{\nonumber}
\newcommand{\C}{\mathbb{C}}
\newcommand{\ovC}{\overline{\C}}
\newcommand{\Csq}{\overline{\C}^{2}}
\newcommand{\Ccsq}{\overline{\C_{c}^{2}}}

\title{Interacting Crumpled Manifolds:\\ Exact Results to all Orders
of Perturbation Theory} 
\author{Henryk A.\ Pinnow\inst{1} \and Kay\ J\"org\
Wiese\inst{2}\footnote{Email: pinnow@theo-phys.uni-essen.de,
wiese@itp.ucsb.edu}} \institute{ \inst{1} Fachbereich Physik,
Universit\"at Essen, 45117
   Essen, Germany\\
  \inst{2} KITP, Kohn Hall, University of California at Santa
Barbara, Santa Barbara, CA 93106, USA}

\begin{document}

\maketitle

\begin{abstract}
In this letter, we report progress on the field theory of polymerized
tethered membranes. For the toy-model of a manifold repelled by a
single point, we are able to sum the perturbation expansion in the
strength $g_{0}$ of the interaction {\em exactly} in the limit of internal
dimension $D\to 2$.  This exact solution is the starting point for an
expansion in $2-D$, which aims at connecting to the well studied case
of polymers ($D=1$). We here give results to order $(2-D)^{4}$, where again
all orders in $g_{0}$ are resummed. This is a first step towards a more
complete solution of the self-avoiding manifold problem, which might
also prove valuable for  polymers. 
\end{abstract}

\section{Introduction} The statistical mechanics of fluctuating lines
and surfaces is a subject of great interest, which poses fundamental
problems and has remained challenging for more than 20 years. One
particular universality class, which has been studied extensively in
the past, are polymerized or ``tethered'' membranes
\cite{JerusalemWinterSchool1989,WieseHabil,DDG1,DDG2,DDG3,DDG4,BowickTravesset2001a,KantorKardarNelson1986b,KantorNelson1987a}. These
are two-dimensional networks, where the bond-length fluctuates, but
never breaks up. In the high-temperature regime nearest-neighbor
interactions can be modeled by a harmonic potential. Neglecting
self-avoidance, the  membrane is extremely  crumpled and highly
folded, a property, which is characterized by the universal
radius-of-gyration exponent $\nu$, defined as
\begin{equation}
  \label{I.0}
  R_{g}\sim L^{\nu}\ , \qquad \nu =0\ ,
\end{equation}
where $R_{g}$ denotes the radius of gyration, and $L$ is the linear
internal size. Physically, $0\leq\nu\leq 1$, but in the absence of
interactions, the radius of 
gyration  grows only logarithmically with the internal size.

For a more realistic description one has to take into account
self-avoidance, whose continuum version  can be  modeled by the generalized
Edwards-Hamiltonian\cite{Edwards1965} with 2-particle contact interaction
\begin{equation}
  \label{I.1}
  \mathcal{H}[\vec r]=\frac{1}{2}\int\limits_{x\in\mathcal{M}}\left(\nabla
  \vec
  r(x)\right)^{2}+\frac{b_{0}}{2}\int\limits_{x\in\mathcal{M}}\int\limits_{y\in\mathcal{M}}\delta^{d}(\vec
  r(x)-\vec r(y))\ ,
\end{equation}
where $x{\in}\mathcal{M}{\subset}\mathbb{R}^{D}$ labels points in the
manifold $\mathcal{M}$, while $\vec r(x){\in}\mathbb{R}^{d}$ points to
their position in external space.  The Edwards model successfully
describes long  polymers
\cite{DesCloizeauxJannink,Schaefer}. Much effort has been spent to
extend these results to membranes ($D=2$).  The problem is, that the
usual $\epsilon$-expansion about the upper critical dimension is not
feasible, since the latter is infinity. An important idea was
therefore to generalize (\ref{I.1}) to manifolds of arbitrary
internal dimension $D$. One then studies the $D$-dimensional manifold
problem, and finally continuous analytically to $D=2$.  A major
breakthrough was the proof of perturbative renormalizability
\cite{DDG3,DDG4} to all orders in perturbation theory.  This procedure
was carried out to two loops \cite{DavidWiese1996,WieseDavid1997}
resulting in a radius-of-gyration exponent of $\nu \approx 0.86$. This
is a strong correction over the non-interacting theory with $\nu=0$,
but may still be in contradiction to Monte-Carlo simulations, which often but
not consistently find tethered membranes in a flat phase with
$\nu=1$
\cite{Baumgaertner1991noNote,BaumgaertnerRenz1992noNote,KrollGompper1993,BowickTravesset2001}.
While simulations are very demanding and therefore not yet conclusive, 
it is nevertheless compelling to try to identify possible mechanisms,
which might render flexible membranes flat at all scales. Such a
mechanism has indeed been found for rigid membranes, where 
fluctuations strongly renormalize rigidity
\cite{NelsonPeliti1987,JerusalemWinterSchool1989}. 

Here we study a simplified model, and solve it
{\em exactly} at $D=2$. It corresponds to a gaussian elastic manifold
interacting by excluded volume with a single $\delta$-like impurity in
external space\cite{Duplantier1989}
\begin{equation}
  \label{I.2}
  \mathcal{H}[\vec r]=\frac{1}{2}\int\limits_{x\in\mathcal{M}}\left(\nabla
  \vec
  r(x)\right)^{2}+g_{0}\int\limits_{x\in\mathcal{M}}\delta^{d}(\vec
  r(x))\ .
\end{equation}
As a first step to prove renormalizability of the full  problem,
\cite{DDG1,DDG2} analysed (\ref{I.2}) and indeed showed
renormalizability to all orders in perturbation theory for all
dimensions  $0<D<2$. (\ref{I.2}) has   essential features
in common with SAM: Its critical embedding dimension tends to infinity
as the internal dimension approaches $D=2$. This can be read off from  the
dimension of the coupling $g_{0}$, which is 
\begin{equation}\label{}
[g_{0}]=:\varepsilon =D -\frac{2-D}{2}d \ .
\end{equation}
Thus, calculating universal
quantities within the $\varepsilon$-expansion necessitates similar
techniques as  for SAM, and we expect to learn more from the solution
of the toy-model (\ref{I.2}).

Recently, we have been able \cite{PinnowWiese2001} to sum the
perturbative expansion {\em exactly} in $D=2$. The key-idea was, that
when approaching $D=2$, the correlator which enters all perturbative
calculations, becomes essentially flat.  In order to check the
consistency of the results obtained by that method, one would like to
go away from $D=2$, and hopefully smoothly connect to polymers in
$D=1$, which are well enough studied to check almost any quantity. In
\cite{PinnowWiese2001}, we have done a first step in that direction,
and obtained quite promising results in first order in
$(2-D)$. However, the expansion in $(2-D)$ is not a loop-expansion,
and at each order in $(2-D)$, we have to resum an infinite number of
diagrams. It turns out, that the results thus become very sensitive to
the regularization procedure.  In this letter, we pursue this road
further, calculating contributions to the partition-function {\em
exactly} for a manifold of toroidal or spherical shape. We obtain the
expansion up to order $(2-D)^{4}$. This information can then be used
to extrapolate away from $D=2$. However, since we find that at $D=2$, the
fixed point is at infinity, one needs
 additional constraints, i.e.\ a scaling function, in order to be able to
use this result. We have not been able to settle this question,
despite the tremendous information contained in the perturbative
result. We thus present our ``raw data'', together with some possible
scaling-functions, encouraging the reader to think himself about the
missing link.

\section{\label{sec:pert}Perturbation theory} Physical observables are
derived from the partition function $\mathcal{Z}(g_{0})$. We use it to
define the effective coupling of the problem,
\begin{equation}
  \label{p.0}
  g(z):=\frac{L^{\varepsilon}}{\mathcal{V}_{\mathcal{M}}}(\mathcal{Z}(0)-\mathcal{Z}(g_{0}))\ ,
\end{equation}
which only depends on the dimensionless combination
$z:=g_{0}L^{\varepsilon}$. $\mathcal{V}_{\mathcal{M}}$ denotes the total
internal volume of the manifold. Accordingly, the perturbation expansion reads
\begin{equation}
  \label{p.1}
g(z)=\frac{g_{0}L^{\varepsilon}}{\mathcal{V_{\mathcal{M}}}}
\sum_{N=0}^{\infty}\frac{(-g_{0})^{N}}{(N{+}1)!}
\left<\prod_{i=1}^{N{+}1}\int_{x_{i}}\tilde\delta^{d}(r(x_{i}))\right>_{0}\
,
\end{equation}
where the normalization of the $\delta$-distribution has been chosen
to be $\tilde\delta^{d}(\vec r(x))=(4\pi)^{d}\delta(\vec r(x))\\
=\int_{k}\mathrm{e}^{i\vec k\vec r(x)}$ with
$\int_{k}:=\pi^{-d/2}\int\mathrm{d}^{d}k$. Performing the averages
within the gaussian theory with normalization $\frac{1}{{\cal V}_{\cal
M}}\int_{x} \left< \tilde\delta^{d}\left( r (x) \right)\right>_{0}=1$,
one arrives at
\begin{equation}
  \label{p.3}
g(z)=\frac{g_{0}L^{\varepsilon}}{\mathcal{V_{\mathcal{M}}}}
\sum_{N=0}^{\infty}\frac{(-g_{0})^{N}}{(N{+}1)!}\left(\prod_{i=1}^{N{+}1}\
\int\limits_{\!  k_{i}}\!\!\int\limits_{x_{i}} \right)
\tilde\delta^{d}\big(\sum_{i}k_{i}\big)\
\mathrm{e}^{\frac{1}{2}\!\!\!\sum
\limits_{i,j=1}^{N{+}1}k_{i}k_{j}C(x_{i}{-}x_{j})}\ ,
\end{equation}
where $C(x):=\frac{1}{2d}\left<(\vec r(x)-\vec r(0))^{2}\right>_{0}$
denotes the  correlator, and the
$\tilde\delta^{d}(\sum_{i}k_{i})$ stems from the integration over the
global translation. Performing the shift $k_{N{+}1}\to
k_{N{+}1}-\sum_{i=1}^{N}k_{i}$ and integrating out the momenta
$k_{1},\dots,k_{N{+}1}$ 
one obtains
\begin{equation}
  \label{p.4} g(z)=z\sum_{N=0}^{\infty}\frac{(-z)^{N}}{(N{+}1)!}
\left(\prod_{\ell=1}^{N}\ \int\limits_{x_{\ell}} \right)(\det
\mathfrak{D})^{-d/2}\ ,
\end{equation}
where we have factored out $L^{\varepsilon}$ from the loop integration
 (such that the integrals now run over a torus of size 1), and
the matrix elements $\mathfrak{D}_{ij}$ are
$\mathfrak{D}_{ij}=\frac{1}{2}[C(x_{N{+}1}{-}x_{i}){+}C(x_{N{+}1}
{-}x_{j}){-}C(x_{i}{-}x_{j})]$.

\section{\label{sec:complete}Complete resummation of the perturbation
series in $D=2$} Let us compute the $N$-loop order of (\ref{p.4}): The
asymptotic behavior of the propagator $C(x)$ for large arguments is of
the form
\begin{equation}
  \label{com.0}
  C(x)\simeq c_{0}+\frac{1}{2\pi}\ln\frac{x}{a}\ ,
\end{equation}
where $c_{0}$ denotes some positive constant (note $C (x)\ge0$), and
the logarithmic growth (for large $x$) is universal.
In $D=2$ we need an additional short distance cutoff $a$. The loop
integrals, denoted by $I_{N}$, only depend on the dimensionless
combination $L/a$. We can (somehow arbitrary) decompose $\det
\mathfrak{D}=(\prod_{i=1}^{N}\mathfrak{D}_{ii})\det \mathfrak{\tilde D}$ with
\begin{eqnarray}\label{com.0.1} \mathfrak{\tilde
D}_{ij}&=&\frac{1}{2}\left[1{+}\frac{C(x_{N{+}1}{-}x_{j})-C(x_{i}{-}x_{j})}
{C(x_{N{+}1}{-}x_{i})}\right]\xrightarrow{a\to
0}\frac{1}{2}\ ,\
    i{\not=}j\ , \nonumber\\
  \mathfrak{\tilde D}_{ii}&=&1\ .
\end{eqnarray}
One has in the limit of $a \to 0$
\begin{equation}\label{com.1} 
\left(\prod_{\ell=1}^{N}\,\int\limits_{x_{\ell}} \right) (\det
\mathfrak{D})^{-d/2}=: I_{N}(L/a)=I_{1}^{N}(L/a)\
(\det\mathfrak{\tilde D}^{(0)})^{-d/2} \ .
\end{equation}
The matrix $ \mathfrak{\tilde D}^{(0)}$ denotes the limit $a\to 0$ of
(\ref{com.0.1}). It can be written as $\mathfrak{\tilde
D}^{(0)}=\frac{1}{2}(\mathbb{I}+N\mathbb{P})$, where $\mathbb{I}$
denotes the identity and $\mathbb{P}$ the projector onto
$(1,1,\dots,1)$, whose image has dimension $1$, such that $\det
\mathfrak{\tilde D}^{(0)}=\frac{1{+}N}{2^{N}}$. Furthermore, to one
loop $I_{1}(L/a)\overset{a\to 0}{=}c_{1}(\ln\frac{L}{a})^{-d/2}$,
where $c_{1}$ denotes some (finite) constant. One then arrives at
\begin{equation}
  \label{com.2}
g(z)=z\sum_{N=0}^{\infty}\frac{(-z(\ln\frac{L}{a})^{-d/2})^{N}}
{N!(1{+}N)^{d/2{+}1}}\
.
\end{equation}
A factor $c_{1}2^{d/2}$ has 
been absorbed into a rescaling of both $z$ and $g$. The above series
can be analysed in the strong coupling limit $z\to \infty$. For this
purpose we define functions $f_{k}^{d}(z)$ together with their
integral representation
\begin{eqnarray} \label{com.3}
f_{k}^{d}(z)&:=&z^{k}\sum_{N=0}^{\infty}\frac{(-z)^{N}}{N!(k{+}N)^{d/2}}=\frac{z^{k}}{\Gamma(\frac{d}{2})}\int\limits_{0}^{\infty}\mathrm{d}r\
    r^{d/2{-}1}\mathrm{e}^{-z\mathrm{e}^{-r}-kr}\nonumber\\
    &=&\frac{(\ln
    z)^{d/2{-}1}}{\Gamma(\frac{d}{2})}\int\limits_{0}^{z}\mathrm{d}y\
    y^{k{-}1}\mathrm{e}^{-y}\left(1-\frac{\ln y}{\ln
z}\right)^{d/2{-}1}
~\xrightarrow{z\to\infty}~
\frac{\Gamma(k)}{\Gamma(\frac{d}{2})}(\ln 
    z)^{d/2{-}1}\ .
  \end{eqnarray}
Thus in the limit of large $z$, the effective coupling (\ref{com.2})
approaches the asymptotic form
\begin{equation}
  \label{com.4} g(z)=\frac{\ \ \big(\!\ln{\textstyle
\frac{L}{a}}\big)^{d/2}}{\Gamma(\frac{d{+}2}{2})}\left[\ln\left(z\left(
\ln{\textstyle \frac{L}{a}}\right)^{-d/2}\right) \right]^{d/2}\ .
\end{equation}

\section{\label{sec:obs}Observables} It immediately follows from this
behavior that the correction-to-scaling exponent $\omega$, which is
defined as the slope of the RG-$\beta$-function at the fixed point,
equals zero. Here, it is useful to study the $\beta$-function as a
function of the bare coupling $z$, which reads
$\beta(z)=-\varepsilon\ z\ \partial g (z) /\partial z$. Then, the
correction-to-scaling exponent is obtained from the limit $z\to\infty$
of 
\begin{equation}\label{omega}
\omega(z):=-\frac{\varepsilon\ z}{\beta(z)} \frac{\partial
\beta (z)}{\partial
  z}\ .
\end{equation}
The value of $\omega$ can be checked in a Monte-Carlo experiment by
considering plaquette-density functions on a membrane with
self-avoidance in only a single $\delta$-like defect.  Be the
partition function $\mathcal{Z}^{\diamond}=\int\mathcal{D}[\vec r]
\tilde \delta^{d}(\vec r(y))\exp[-\mathcal{H}[\vec r]]$, then the
plaquette-density at the defect is obtained from
$\left<n\right>_{\diamond}=\frac{L^{\varepsilon}}{\partial g/\partial
z}\frac{\partial}{\partial z}(\frac{\partial g}{\partial z})$, where
$\frac{\partial g}{\partial z}=\mathcal{Z}^{\diamond}$. One
furthermore needs the density-density correlation at this point, which
is defined as
$\left<n^{2}\right>_{\diamond}=\frac{L^{2\varepsilon}}{\partial
g/\partial z}\frac{\partial^{2}}{\partial z^{2}}(\frac{\partial
g}{\partial z})$. In the limit of strong coupling
$\left<n\right>_{\diamond}=\frac{1}{g_{0}}(1{+}\frac{\omega}{\varepsilon})$
and $\left<n^{2}\right>_{\diamond}=\frac{1}{g_{0}^{2}}
(2{+}\frac{3\omega}{\varepsilon}{+}\frac{\omega^{2}}{\varepsilon^{2}})$,
such that the ratio \begin{equation} \label{obs.0}
\frac{\left<n\right>_{\diamond}}
{\sqrt{\left<n^{2}\right>_{\diamond}}}~\xrightarrow{z\to\infty}~
\sqrt{\frac{\varepsilon{+}\omega}{2\varepsilon{+}\omega}}
~\xrightarrow{\omega=0}~\sqrt{\frac{1}{2}}
\end{equation} becomes universal and should be  measurable in simulations.

\section{\label{sec:2mD}$( 2-D)$-expansion} Let us now analyse the
theory below $D=2$. Due to the renormalizability in $0<D<2$ and the
existence of an $\varepsilon$-expansion we expect the renormalized
coupling to reach a finite fixed point in the strong coupling limit as
soon as $D<2$. This approach is characterized by a powerlaw decay of
the form
\begin{equation}\label{2mD.0}
  g(z)=g^{*}+S(\ln z)\ z^{-\omega/\varepsilon}+
O(z^{-\omega_{1}/\varepsilon})\ ,
\end{equation}
where $S$ is some scaling-function growing at most sub-exponentially
and $\omega_{1} >\omega >0$, with $\omega$ defined in
(\ref{omega}). In order to gain information about $g$ below $D=2$ one
has to expand the loop integrand $(\det \mathfrak{D})^{-d/2}$ in powers of
$2{-}D$. For convenience, we take $a\to 0$. The propagator
(\ref{com.0}) takes in infinite $D$-space the form $C(x)=|x|^{2{-}D}/(S_{D}(2{-}D))$, where
$S_{D}=2\pi^{D/2}/\Gamma(\frac{D}{2})$ denotes the volume of the
$D$-dimensional unit-sphere. The factor $(S_{D}(2{-}D))^{-1}$ replaces
$\ln (\frac{L}{a})$ and is absorbed into a rescaling of the field and
the coupling according to $r\to r\ (S_{D}\ (2{-}D))$ and $g_{0}\to
g_{0}\ (S_{D}\ (2{-}D))^{d/2}$, such that the factors of
$(\ln\frac{L}{a})^{-d/2}$ in (\ref{com.2}) and (\ref{com.4}) are
replaced by $(S_{D}\ (2{-}D))^{d/2}$. The propagator in the rescaled
variable can then be written as
\begin{equation}
  \label{2mD.1}
  C(x)=1+(2-D)\ \mathbb{C}(x)\ .
\end{equation}
where for convenience of notation we allow $\mathbb{C} (x)$ to depend
itself on $D$.

Of course, on a closed manifold of finite size, $C(x)$ needs to be
modified, but the form (\ref{2mD.1}) is independent of the shape of
the manifold. Accordingly, one may expand the matrix $\mathfrak{D}$, which
is $\mathfrak{D}=\mathfrak{\tilde D}^{(0)}+(2{-}D)\ \mathbb{D}$, where
$\mathfrak{\tilde D}^{(0)}$ is defined as before and coincides with the
limit $D{\to}2$ when inserting the above $C(x)$ into
$\mathfrak{D}$. Moreover, $\mathbb{D}$ is of the same form as $\mathfrak{D}$,
but each $C (x)$ has been replaced with $\mathbb{C}(x)$:
$\mathbb{D}_{ij}=\frac12 [\mathbb{C} (x_{N+1}-x_{i})+\mathbb{C}
(x_{N+1}-x_{j})+\mathbb{C} (x_{i}-x_{j}) ]$. Then,
\begin{equation}\label{2mD.2}
\det \mathfrak{D}=\det 
\mathfrak{\tilde D}^{(0)}\, \exp\left\{ \mathrm{Tr}\left[  \ln(1+(2-D)[
\mathfrak{\tilde D}^{(0)}]^{-1}\mathbb{D}) \right] \right\}\ ,
\end{equation}
where $[ \mathfrak{\tilde
D}^{(0)}]^{-1}=2(\mathbb{I}{-}\frac{N}{N{+}1}\mathbb{P})$ denotes the
inverse matrix of $\mathfrak{\tilde D}^{(0)}$. Expanding the integrand
(\ref{2mD.2}) in powers of $(2-D)$ and the coupling $g_{0}$, all
orders in $g_{0}$ can again be summed, with the difference that the
integrands are no longer constant. Expanding up to the $n$th order in
$2{-}D$ involves $n$ powers of $\mathbb{C}(x)$. Introducing the
notation $\overline{f(x_{{1}},\dots,x_{{k}})} :=
\int_{x_{1}}\cdots\int_{x_{k}}f(x_{{1}},\dots,x_{{k}})$ with the
integration defined as $\int_{x}:=\int\mathrm{d}^{D}x $ (on the torus)
the overbar can be thought of as an averaging procedure. To first and
second order in $2{-}D$, the only integrals to be evaluated are
$\overline{\mathbb{C}(x)}$ and $\overline{\mathbb{C}^{2}(x)}$. In
order to reveal the structure of the expansion we generated all terms
up to fourth order. Generally, the terms are of the following form
\begin{eqnarray}
  \label{2mD.2p} z\sum_{N=1}^{\infty}(\det \mathfrak{\tilde
D}^{(0)})^{-d/2}\, \frac{\overline{\prod_{i=1}^{l}\left(\Tr ([
\mathfrak{\tilde
D}^{(0)}]^{-1}\mathbb{D})^{n_{i}}\right)^{m_{i}}}\,(-z)^{N}}{(N{+}1)!}
&=:& \sum_{j=\mbox{\tiny{min}}}^{\mbox{\tiny{max}}}\mathbb{M}_{
\tiny{\left(\begin{array}{cccc}
        \!\!m_{1} & \!\!\!\!m_{2} & \!\!\!\!\cdots & \!\!\!\!m_{l}\\
        \!\!n_{1} & \!\!\!\!n_{2} & \!\!\!\!\cdots & \!\!\!\!n_{l}
    \end{array}\!\!\right)}
}f_{1}^{d+2j}(z)\nonumber\\
  &=: &\mathbb{M}^{j}_{
    \tiny{\left(\begin{array}{cccc}
        \!\!m_{1} & \!\!\!\!m_{2} & \!\!\!\!\cdots & \!\!\!\!m_{l}\\
        \!\!n_{1} & \!\!\!\!n_{2} & \!\!\!\!\cdots & \!\!\!\!n_{l}
    \end{array}\!\!\right|\left.\begin{array}{c}
      \!\! \mbox{max}\\
      \!\! \mbox{min}
    \end{array}\right)}
}f_{1}^{d+2j}(z)\ ,\qquad 
\end{eqnarray}
where $\mbox{\small{max}}$ and $\mbox{\small{min}}$ are some integers,
and summation over the index $j$ is implicit. The precise form of the
vector entries $\mathbb{M}^{j}$ will be reported
elsewhere \cite{PinnowWieseInPreparation}. The renormalized coupling then
reads up to  fourth order in $2{-}D$ (note that we have absorbed a
factor of $2^{d/2}$ in both $g$ and $z$):
\begin{eqnarray}
  \label{2mD.3}
  g(z)&=&f_{1}^{d+2}(z)-(2{-}D)\
\frac{d}{2}\mathbb{M}^{j}_{(^{1}_{1}|^{1}_{0})}f_{1}^{d+2j}(z)\nn\\ 
  &&+(2{-}D)^{2}\left[\frac{d}{4}\mathbb{M}^{j}_{
    {(^1_2|^{~2}_{{-1}})}
}f_{1}^{d+2j}(z)+\frac{d^{2}}{8}\mathbb{M}^{j}_{
    {(^{2}_{1}|^{~2}_{-1})}
}f_{1}^{d+2j}(z)\right]\nn\\
  &&-(2{-}D)^{3}\left[\frac{d}{4}\mathbb{M}^{j}_{
    {(^{1}_{3}|^{~2}_{-3})}
}f_{1}^{d+2j}(z)+\frac{d^{2}}{8}\mathbb{M}^{j}_{
    {(^{1}_{1}~^{1}_{2}|^{~3}_{-2})}
}f_{1}^{d+2j}(z)+\frac{d^{3}}{48}\mathbb{M}^{j}_{
    {(^{3}_{1}|^{~3}_{-2})}
}f_{1}^{d+2j}(z)\right]\nn\\
  &&+(2{-}D)^{4}\left[\frac{d}{8}\mathbb{M}^{j}_{
    {(^{4}_{1}|^{~4}_{{-3}})}
}f_{1}^{d+2j}(z)+\frac{d^{2}}{8}\left(\frac{1}{4}\mathbb{M}^{j}_{
    {(^{2}_{2}|^{~4}_{-3})}
}f_{1}^{d+2j}(z)+\frac{2}{3}\mathbb{M}^{j}_{
    {(^{1}_{1}~^{1}_{3}|^{~4}_{-3})}
}f_{1}^{d+2j}(z)\right)\right.\nn\\
  &&\hspace{1.8cm}\left.+\frac{d^{3}}{32}\mathbb{M}^{j}_{
    {(^{2}_{1}~^{1}_{2}|^{~4}_{-3})}
}f_{1}^{d+2j}(z)+\frac{d^{4}}{384}\mathbb{M}^{j}_{
    {(^{4}_{1}|^{~4}_{-3})}
}f_{1}^{d+2j}(z)\right]\ +\ O(2-D)^{5}
\end{eqnarray}
From the integral representation (\ref{com.3}) of $f_{1}^{d{+}j}(z)$
and the above expansion, it follows immediately that the exact
renormalized coupling can be written as
\begin{equation}\label{2mD.4} 
g(z)=z\int_{0}^{\infty}\mathrm{d}r\ \tilde g ( r)\
\mathrm{e}^{-z\mathrm{e}^{-r}-r}\ ,
\end{equation}
where $\tilde g(r)$ is of the form
\begin{equation}\label{2mD.5} 
\tilde g(r)=r^{d/2}\bigg[\frac{1}{\Gamma(\frac{d{+}2}{2})}+
(2{-}D)\sum\limits_{n=0}^{\infty}\,\sum
\limits_{j=-n_{\mathrm{max}}}^{n} p_{n_{j}}r^{j}(2{-}D)^{n}\bigg]
\ .
\end{equation}
Let us try to gain more information about the powerlaw behavior in
(\ref{2mD.0}), that is about the expansion in $2{-}D$ of the
correction-to-scaling exponent $\omega$. Powerlaw behavior forces the
series (\ref{2mD.5}) to turn into some exponentially decaying
function $\tilde g(r)$ as can be seen from
the asymptotic form of $g (z)$ 
\begin{equation}\label{2mD.6}
 g (z)\simeq {\cal A} + {\cal B} z^{-\omega/\E}= {z}
\int\limits_{0}^{\infty}\mbox{d}r\ \mathrm{e}^{-z\ \rme^{-r}-r}
\left({\mathcal{A}}+\frac{\mathcal{B}\,\rme^{-r \omega /\epsilon }}
{{\Gamma(1{+}\frac{\omega}{\varepsilon})}} \right) +O(\mathrm{e}^{-z})
\end{equation}
Now, we test a possible form of the exact $\tilde g(r)$, which
is consistent with the expansion (\ref{2mD.3}) and which satisfies the
following properties: (i) In the limit of $D=2$ the exact form
$r^{d/2}/\Gamma(\frac{d{+}2}{2})$ emerges and (ii) for $D<2$ the
corresponding $g(z)$ has a finite fixed-point value together with a
strong coupling expansion. The (non-unique) ansatz is
\begin{equation}
  \label{2mD.7}
  \tilde g(r)=\mathcal{C}\left(\frac{1-{\mathcal S}(D,r)\
  \mbox{e}^{-\frac{\omega}{\E}r}}{\omega/\E}\right)^{d/2}\ ,
\end{equation}
where ${\mathcal S} (D,r)$ is analytic in $D=2$ of the form $
{\mathcal S}(D,r)=1+\frac{\omega}{\epsilon}r
\sum_{n=1}^{\infty}{\mathcal S}_{n}(r)(2{-}D)^{n}$, and each
${\mathcal S}_{n}(r)$ has a Laurent expansion $ {\mathcal
S}_{n}(r)=\sum_{j=-n_{\tiny{\mbox{min}}}}^{n_{\tiny{\mbox{max}}}}
s_{n,j}\, r^{j}$. Note, that in the limit of $D\to 2$, the expression
 (\ref{2mD.7}) gives $r^{d/2}$, while for $D<2$ it yields upon
integration the form (\ref{2mD.6}), ensuring both properties (i) and
(ii). Inserting
$\omega/\varepsilon=\omega_{2}(2{-}D)^{2}{+}O(2{-}D)^{3}$ (the linear
term in $(2{-}D)$ has to vanish) into the ansatz (\ref{2mD.7}) and
expanding to second order in $2{-}D$ provides
\begin{equation}\label{2mD.8}
 \tilde g(r)=\mathcal{C}\
r^{d/2}\left[1{-}\frac{d}{2}\left({\mathcal
S}_{1}(r)(2{-}D)+\left(\frac{\omega_{2}}{2}r-\frac{d-2}{4}{\mathcal
S}_{1} (r)^{2} 
+ {\mathcal S}_{2}(r)\right)(2{-}D)^{2}{+}\cdots\right)\right]\ .
\end{equation}
The first coefficients of the  $( 2{-}D)$-expansion
of $\tilde g(r)$ obtained from (\ref{2mD.3}) read
\begin{eqnarray}
    \label{2mD.9} 
\tilde g(r) \!&=&\! \frac{r^{d/2}}{\Gamma (\frac{d{+}2}{2})}
\left\{ 1+ (2{-}D)\frac{d}{2}\ovC\left(1 {-}\frac{d}{2r}\right)\right.\!
- (2{-}D)^{2}\!
\left[\frac{d}{2}\Ccsq\ r{+}\frac{d}{4}\left(\Csq{-}4\Ccsq\right){-}\frac{d^{2}}{8}\left(2\Ccsq{+}\Csq\right)\right.\!\!\!\nn\\
&&\hphantom{\frac{r^{d/2}}{\Gamma (\frac{d}{2}+1)}[}\left.\left.
{+}\!\left(\frac{d^{2}}{8}\left({-}\Csq{+}3\Ccsq\right){+}\frac{d^{3}}{8}\Csq\right)r^{{-}1}{-}\frac{d^{2}}{8}\left(\frac{d}{2}{-}1\right)\left(\Ccsq{+}\frac{d}{2}\Csq\right)r^{{-}2}\right]\right\}\ .
\end{eqnarray}
Comparing    (\ref{2mD.8})   and    (\ref{2mD.9}),    one   identifies
$\mathcal{C}=1/\Gamma(\frac{d{+}2}{2})$,          $          {\mathcal
S}_{1}=-\overline{\mathbb{C}}(1{-}\frac{d}{2}\frac{1}{r})$          and
$\omega_{2}=2\overline{\mathbb{C}_{c}^{2}}$,                      where
$\mathbb{C}_{c}(x){:=}\mathbb{C}(x)-\overline{\mathbb{C}}$.  Note that
the  terms proportional  to $\overline{C}^{2}$  in ${\cal  S}_{2} (r)$
mostly cancel with ${\cal S}_{1} (r)^{2}$,  a sign that the ansatz catches
some  structure.

The  diagrams  to  be calculated  at  this order  are
$\overline{\C}$ and $\overline{\C_{c}^{2}}$. On a manifold of toroidal
shape,  which  is  equivalent  to periodic  boundary  conditions,  two
discrete sums have to be evaluated:
\begin{eqnarray}\label{2mD.10} 
\overline{\C}&=&S_{D}\bigg[ \sum_{\vec k\not=\vec 0}\frac{1}{{\vec
k}^{2}}-\frac{1}{2\pi(2-D)}\bigg]=-0.44956 +
0.3583\, (2-D)+  O(2-D)^{2}\qquad \\   
  \label{2mD.11}
\overline{\C_{c}^{2}}&=&S_{D}^{2}\sum_{k\not=0}\frac{1}{{\vec
k}^{4}}=0.152661 + O(2-D)\ .
\end{eqnarray}
$\vec k$ is $D$-dimensional with components $k_{i}=2\pi/L\ n_{i}$, and
$n_{i}$ integer. With the results given above, this leads to
\begin{equation}\label{finalomega}
\omega = 2 \epsilon \overline{\C_{c}^{2}} (2-D)^{2} + O (2-D)^{3} =
0.305322\, \epsilon\,(2- D)^{2} + O (2-D)^{3}\ ,  
\end{equation}
which can be compared to the exact result for $D=1$ (polymers):
$\omega =\epsilon$.  As a caveat, note that the above scheme is not
unambiguous, since different ans\"atze in (\ref{2mD.7}) are possible.
Also the second order term proportional to $r$ in (\ref{2mD.9}) could
in principle either be attributed to $\omega_{2}$ or ${\cal
S}_{2}$. More constraints are necessary to settle this question.

In summary: We have presented a complementary approach to treat the
problem of tethered membranes in interaction. We hope that this
approach will prove fruitful for self-avoiding tethered membranes,
with eventual applications  for polymers.

\acknowledgments
It is a pleasure to thank H.W.\ Diehl, A.\ Ludwig  and L.\ Sch\"afer  for stimulating discussions. This work has  been supported  
by the DFG through the Leibniz program Di 378/2-1, under Heisenberg
grant Wi 1932/1-1, and by the  NSF under grant  PHY99-07949.


\end{document}